\documentclass[12pt,a4,pdftex,revtex,amsfonts,amsmath,amssymb]{article}
\usepackage{amsmath}
\usepackage{graphicx}
\usepackage{color}

\begin{document}


\begin{center}
\large{Hydration of Clays at the Molecular Scale: The Promising Perspective of Classical Density Functional Theory}\\
\vspace{8mm}
 \small{ \textbf{ Guillaume Jeanmairet$^{a,b,}$\footnote{guillaume.jeanmairet@ens.fr}, Virginie Marry$^{c,d}$, Maximilien Levesque$^{a,b,}$\footnote{maximilien.levesque@ens.fr}, Benjamin Rotenberg$^{c,d}$, Daniel Borgis$^{a,b}$}}
\\\vspace{6pt} $^{a}${\em{\'Ecole Normale Sup\'erieure-PSL Research University, D\'epartement de Chimie, 24, rue Lhomond, 75005 Paris, France b: CNRS, UMR 8640 PASTEUR, F-75005, Paris, France. c: Sorbonne UniversitŽs, UPMC Univ Paris 06, PASTEUR, F-75005, Paris, France.}};
\\\vspace{6pt} $^{b}${\em{CNRS UMR 8640 PASTEUR, 75005 Paris, France}};
\\\vspace{6pt} $^{c}${\em{Sorbonne Universit\'es, UPMC Univ Paris 06, UMR 8234 PHENIX, F-75005, Paris, France}};
\\\vspace{6pt} $^{d}${\em{CNRS, UMR 8234 PHENIX, F-75005, Paris, France}};
\\\vspace{6pt}
\end{center}

\begin{abstract}
We report here how the hydration of complex surfaces can be efficiently
studied thanks to recent advances in classical molecular density functional theory.
This is illustrated on the example of the pyrophylite clay.
After presenting the most recent advances, we show that the
strength of this implicit method is that (\emph{i}) it is in quantitative
or semi-quantitative agreement with reference all-atoms simulations
(molecular dynamics here) for both the solvation structure and energetics,
and that (\emph{ii}) the computational cost is two to three orders
of magnitude less than in explicit methods. The method remains imperfect,
in that it locally overestimates the polarization of water close to
hydrophylic sites of the clay. The high numerical efficiency of the method
is illustrated and exploited to carry a systematic study of the electrostatic
and van der Waals components of the surface-solvant interactions
within the most popular force field for clays, CLAYFF. Hydration structure
and energetics are found to weakly depend upon the electrostatics.
We conclude on the consequences of such findings on future force-field
development.
\end{abstract}

\section{Introduction}

Pierre Turq, in his recent conference ``Histoire d'Eau'', at Universit\'e
Pierre et Marie Curie, Paris, highlighted the immense challenge that
represents modeling of water at the atomic scale in complex, hierarchical
materials such as clays\cite{meunier_argiles_2003}. Their surface to
volume ratio is high, and have thus key role in various applications
such as heterogenous catalysis, electrochemistry, adsorption and transport
in porous media. In such materials, being able to model at
the same time the smallest and the largest representative scales, \textit{e.g.},
keeping the molecular picture of the solvent while studying mesoscale
transport, remains an immense open field. It is our goal to develop tools
that will improve the description at the molecular scale, while keeping
open the upscaling.

Experimental tools have been developed to get insight into the properties
of water at clay surfaces: vibrational sum frequency spectroscopy\cite{richmond_molecular_2002},
quasi-elastic neutron scattering techniques\cite{marry_water_2011},
infrared (IR) and Raman spectroscopies or nuclear magnetic resonance
(NMR). Atomic scale modeling can clarify or give insight into
these experimental observations\cite{marry_water_2011,botan_hydrodynamics_2011,marry_structure_2008,marry_anisotropy_2013}.
It was for instance the case of hydrophobic versus hydrophilic behaviours
of various clay surfaces and humidities\cite{michot_structural_1994,van_oss_hydrophilicity_1995} that has been recently clarified
by molecular dynamics simulations and theoretical insights\cite{rotenberg_molecular_2011}.
Water dynamics in hectorite clay have also been studied by coupled neutron
spin echo and molecular dynamics\cite{marry_water_2011}.

This kind of theoretical modeling most often relies on molecular
dynamics (MD) and Monte Carlo (MC) simulations, in which each atom is considered
individually. Observables are statistical averages over configurations that are obtained deterministically or stochastically. With
sufficient sampling, they are theoretically exact. Nevertheless, they
are computationally demanding. Nowadays, multiscale materials with several (thousands of) thousands of atoms can be simulated by MD or MC but require using
supercomputers and thousands of threads of computational processing
units (CPU). At this point, numerical heaviness becomes critical and
hinders systematic studies. 

To overcome this numerical difficulty, coupled to energetic and economic
considerations of running supercomputers, various empirical
modeling and physically grounded liquid state theories have emerged.
One can for instance forget about the molecular nature of the solvent
to only account for a polarizable continuum model (PCM)\cite{PCM_leszczynski_computational_1999,PCM_tomasi99,PCM_Tomasi05_review_continuummodels},
or simplify the solvent molecule to a dipolar hard sphere particle\cite{DHS_biben_generic_1998,DHS_dzubiella_competition_2004}.
Coarse grained strategies have also been successfully applied to solid-liquid interfaces\cite{jardat_salt_2009,jardat_self-diffusion_2012}.
Nevertheless, while all these methods proved  their efficiency on a panel
of problems, the applicability and transferability of their underlying
approximations remain to be systematically checked.

These last forty years, liquid state theories\cite{gray_theory_1984}
have blossomed, classical density functional theory (DFT)\cite{evans79,evans92,wu_density-functional_2007}, integral equations, classical field
theories, all have shown to be able to give solvation properties in
at least qualitative agreement with reference MD or MC simulations.
Nevertheless, they remained toy theories for a long time, \textit{i.e.},
applied only to problems of high symmetry like hard and soft walls.
A current challenge lies in the development of a three-dimensional
liquid state theory (and its implementation) for complex liquids to
describe molecular liquids, solutions, mixtures, in 
complex environments such as biomolecular media or atomistically resolved
surfaces and interfaces. Recent developments in this direction have
been done within lattice field theory\cite{lattice_field_azuara_pdb_hydro:_2006,lattice_field_azuara_incorporating_2008},
Gaussian field theory\cite{gaussian_field_varilly_improved_2011},
three-dimensional reference interaction site model (3D-RISM)\cite{3D_RISM_beglov_integral_1997,3D_RISM_Hirata_Molecular_Theory_of_Solvation,3D_RISM_kloss_quantum_2008,3D_RISM_kloss_treatment_2008,3D_RISM_kovalenko_three-dimensional_1998,3D_RISM_yoshida_molecular_2009}.
This last integral equation theory has proved to be useful
to a large panel of applications like structure prediction in complex
biomolecular systems, but remains difficult to control and improve
because of the choice of a closure relation, typically restricted
to Hypernetted chain (HNC), Percus-Yevick\cite{percus_analysis_1958},
or Kovalenko-Hirata\cite{3D_RISM_kovalenko_three-dimensional_1998}.

Molecular density functional theory (MDFT) has recently been shown
to be a powerful tool to study the solvation of complex surfaces
and interfaces by a polar solvent, at a fully molecular level of description\cite{MDFT_ramirez_density_2002,MDFT_ramirez_density_2005,MDFT_ramirez_direct_2005,MDFT_gendre_classical_2009,MDFT_zhao_molecular_2011,MDFT_borgis_molecular_2012,jeanmairet_molecular_2013-1,jeanmairet_molecular_2013}.
In the case of the generic model of dipolar solvent, \textit{i.e.}, the Stockmayer
fluid (eventually parameterized to mimic a few properties of water),
it was shown to recover quantitatively the structural solvation
properties of a complex atomically resolved clay surface, the pyrophyllite, that contains over a thousand atoms\cite{levesque_solvation_2012}.
 It is the objective of this article to test the most recent developments of
MDFT for one of the most important but challenging fluid, namely water, in the typical complex case of the pyrophyllite clay.

In Sec.~\ref{sec:Theory-and-Methods}, we present the theoretical and numerical recent
developments of MDFT for water.
We also briefly introduce our molecular dynamics simulations, that
will be used as a reference. In Sec.~\ref{sec:Results}, we apply the method to the
solvation of a pyrophyllite clay sheet by water. We investigate both
structural and energetic aspects of this hydration. Finally, in Sec.~\ref{sec:Conclusion}
, we conclude and give perspectives to the method to overcome actual
limitations.

\section{Theory and Methods\label{sec:Theory-and-Methods}}

\subsection{Theory}

Molecular density functional theory relies on the definition of a
free-energy functional of the six-dimensional position and orientation
molecular one-particle density, $\rho\left(\mathbf{r},\boldsymbol{\Omega}\right)$,
where $\boldsymbol{\Omega}$ accounts for the 3 Euler angles. This
functional remains unknown and various approximations have been proposed.
In the homogeneous reference fluid approximation (HRF), the functional
is built from the properties of the bulk solvent. This approach gives
accurate results for polar, aprotic fluids\cite{MDFT_gendre_classical_2009,MDFT_zhao_molecular_2011,MDFT_borgis_molecular_2012}.
For its part, water requires special attention\cite{jeanmairet_molecular_2013,jeanmairet_molecular_2013-1,MDFT_levesque_krfmt}. We recall here some of the development made to treat water; note that they are also adapted to any solvent with only one Lennard-Jones site.
Until now, only simple three-dimensional solutes have been tested in this
MDFT framework.
Let us define several microscopic densities of a system of $N$ particles: the position and orientation
density of solvent molecules, $\rho\left(\mathbf{r},\boldsymbol{\Omega}\right)$,
the position density, $n\left(\mathbf{r}\right)$, the charge density,
$\rho_{c}\left(\mathbf{r}\right)$, and the \emph{multipolar} polarization
density field, $\boldsymbol{\mathbf{P}}\left(\mathbf{r}\right)$:
\begin{eqnarray}
\rho\left(\mathbf{r},\boldsymbol{\Omega}\right) & = & \left\langle \sum_{i=1}^{N}\delta\left(\mathbf{r}-\mathbf{r}_{i}\right)\delta\left(\boldsymbol{\Omega}-\boldsymbol{\Omega}_{i}\right)\right\rangle ,\label{eq:rho_solvant}\\
n\left(\mathbf{r}\right) & = & \left\langle \sum_{i=1}^{N}\delta\left(\mathbf{r}-\mathbf{r}_{i}\right)\right\rangle =\int\rho\left(\mathbf{r},\boldsymbol{\Omega}\right)\text{d}\boldsymbol{\Omega},\label{eq:n}\\
\rho_{c}\left(\mathbf{r}\right) & = & \left\langle \sum_{i=1}^{N}\sigma\left(\mathbf{r}-\mathbf{r}_{i},\boldsymbol{\Omega}_{i}\right)\right\rangle ,\\
 & = & \iint\sigma\left(\mathbf{r}-\mathbf{r}^{\prime},\boldsymbol{\Omega}\right)\rho\left(\mathbf{r}^{\prime},\boldsymbol{\Omega}\right)\text{d}\mathbf{r}^{\prime}\text{d}\boldsymbol{\Omega},\\
\boldsymbol{\mathbf{P}}\left(\mathbf{r}\right) & = & \left\langle \sum_{i=1}^{N}\boldsymbol{\mu}\left(\mathbf{r}-\mathbf{r}_{i},\boldsymbol{\Omega}_{i}\right)\right\rangle ,\label{eq:P}\\
 & = & \iint\boldsymbol{\mu}\left(\mathbf{r}-\mathbf{r}^{\prime},\boldsymbol{\Omega}_{}\right)\rho\left(\mathbf{r}^{\prime},\boldsymbol{\Omega}\right)\text{d}\mathbf{r}^{\prime}\text{d}\boldsymbol{\Omega},
\end{eqnarray}
where $\langle \dots \rangle$ indicates an ensemble average and $\text{d}\boldsymbol{\Omega}$ an angular integration
over the three Euler angles, \textit{i.e.}, over all possible orientations of the water molecules and all possible rotations
 around their $C_2$ axis. The molecular charge density, $\sigma\left(\mathbf{r},\boldsymbol{\Omega}\right)$,
and the molecular polarization density field of a water molecule at the
origin, $\boldsymbol{\mu}\left(\mathbf{r},\boldsymbol{\Omega}\right)$,
are:
\begin{eqnarray}
\sigma\left(\mathbf{r},\boldsymbol{\Omega}\right) & = & \sum_{m}q_{m}\delta\left(\mathbf{r}-\mathbf{s}_{m}\left(\boldsymbol{\Omega}\right)\right),\\
\boldsymbol{\mu}\left(\mathbf{r},\boldsymbol{\Omega}\right) & = & \sum_{m}q_{m}\mathbf{s}_{m}\left(\boldsymbol{\Omega}\right)\int_{0}^{1}\delta\left(\mathbf{r}-u\mathbf{s}_{m}\left(\boldsymbol{\Omega}\right)\right)\text{d}u,
\end{eqnarray}
where $\mathbf{s}_{m}\left(\boldsymbol{\mathbf{\Omega}}\right)$ is
the position of the $m^{\text{th}}$ (between 1 and 3) atomic site
of the water molecule with orientation $\boldsymbol{\Omega}$. The
molecular charge density and polarization density field are linked
by the relation $\sigma=-\nabla\cdot \boldsymbol{\mu}$. At first order,
the multipolar polarization density field reduces to a molecular dipole
located at the origin.

For its part, the solute is described by sites, $i$, located at $\mathbf{R}_{i}$,
with Lennard-Jones parameters $\epsilon_{i}$ and $\sigma_{i}$, and
point charges $q_{i}$. The force field associated of the solute is
given in Tab.~\ref{tab:Description-of-the}. We use the Lorentz-Berthelot
mixing rules to determine $\sigma_{ij}$ and $\epsilon_{ij}$. The
solute induces a scalar external potential, $\phi_{n}$, and an electric
field, $\mathbf{E}_{c}$:
\begin{eqnarray}
\phi_{n}\left(\mathbf{r}\right) & = & \sum_{j}4\epsilon_{ij}\left[\left(\frac{\sigma_{ij}}{\left|\mathbf{r}-\mathbf{R}_{j}\right|}\right)^{12}-\left(\frac{\sigma_{ij}}{\left|\mathbf{r}-\mathbf{R}_{j}\right|}\right)^{6}\right],\\
\mathbf{E}_{c}\left(\mathbf{r}\right) & = & \frac{1}{4\pi\epsilon_{0}}\sum_{j}q_{j}\frac{\mathbf{r}-\mathbf{R}_{j}}{\left|\mathbf{r}-\mathbf{R}_{j}\right|^{3}},
\end{eqnarray}
where $\epsilon_{0}$ is the vacuum permitivity.

We can extend the original derivation by Evans\cite{evans79,evans92}
of the molecular density functional theory to define the solvation free energy functional,  $\mathcal{F}$, depending of position density and the polarization density.
$\cal F$ is expressed as the difference in grand potential, $\Theta-\Theta_{b}$, of the
system composed of the solute plus the solvent at a given chemical
potential, and of the sole homogeneous reference water fluid at bulk
density, $\rho_{b}=n_b/(8\pi^2)$, with $n_b\approx$0.033 molecules per \AA$^{3}$:
\begin{eqnarray} \label{eq:FF} 
\mathcal{F}\left[n\left(\mathbf{r}\right),\mathbf{P}\left(\mathbf{r}\right)\right]&=&\Theta\left[n\left(\mathbf{r}\right),\mathbf{P}\left(\mathbf{r}\right)\right]-\Theta_{b}\\
&=&\mathcal{F}_{\text{int}}\left[n\left(\mathbf{r}\right),\mathbf{P}\left(\mathbf{r}\right)\right]+\int\phi\left(\mathbf{r}\right)n\left(\mathbf{r}\right)\text{d}\mathbf{r}
-\int\mathbf{E}_{c}\left(\mathbf{r}\right)\cdot\mathbf{P}\left(\mathbf{r}\right)\text{d}\mathbf{r},\nonumber 
\end{eqnarray}
where the intrinsic Helmholtz free energy functional, $\mathcal{F}_{\text{int}}\left[n\left(\mathbf{r}\right),\mathbf{P}\left(\mathbf{r}\right)\right]$,
can be decomposed into the sum of an ideal part and an excess part,
$\mathcal{F=\mathcal{F}_{\text{id}}}+\mathcal{F}_{\text{exc}}$. The
ideal part is known rigorously:
\begin{eqnarray} \label{eq:Fid}
\beta\mathcal{F}_{\text{id}}\left[n\left(\mathbf{r}\right),\mathbf{P}\left(\mathbf{r}\right)\right] & = & \iint\big[\rho\left(\mathbf{r},\boldsymbol{\Omega}\right)\ln\left(\frac{\rho\left(\mathbf{r},\boldsymbol{\Omega}\right)}{\rho_{b}}\right)
  -\rho\left(\mathbf{r},\boldsymbol{\Omega}\right)+\rho_{b}\big]\text{d}\mathbf{r}\text{d}\boldsymbol{\Omega} 
\end{eqnarray}
where $\beta\equiv\left(k_{B}T\right)^{-1}$,
$k_{B}$ is the Boltzmann constant, and $T$ the absolute temperature.
On its side, the exact excess term is unknown. Finding a good term
for $\mathcal{F}_{\text{exc}}$ is a current challenge. We proposed
recently to expand it around the homogeneous liquid state at the same
density, $\rho_{b}$, where we use as sole input the experimental
or calculated structure factor, $S$, and the experimental longitudinal
and transverse electric susceptibilities, $\chi_{L}$ and $\chi_{T}$. The reader is referred to Ref.~\cite{jeanmairet_molecular_2013-1}
for the complete expression of the excess term of the Helmholtz free
energy functional we use therein.

\subsection{System Description\label{Sys_def}}

\begin{figure}
\begin{centering}
\includegraphics[width=1\columnwidth]{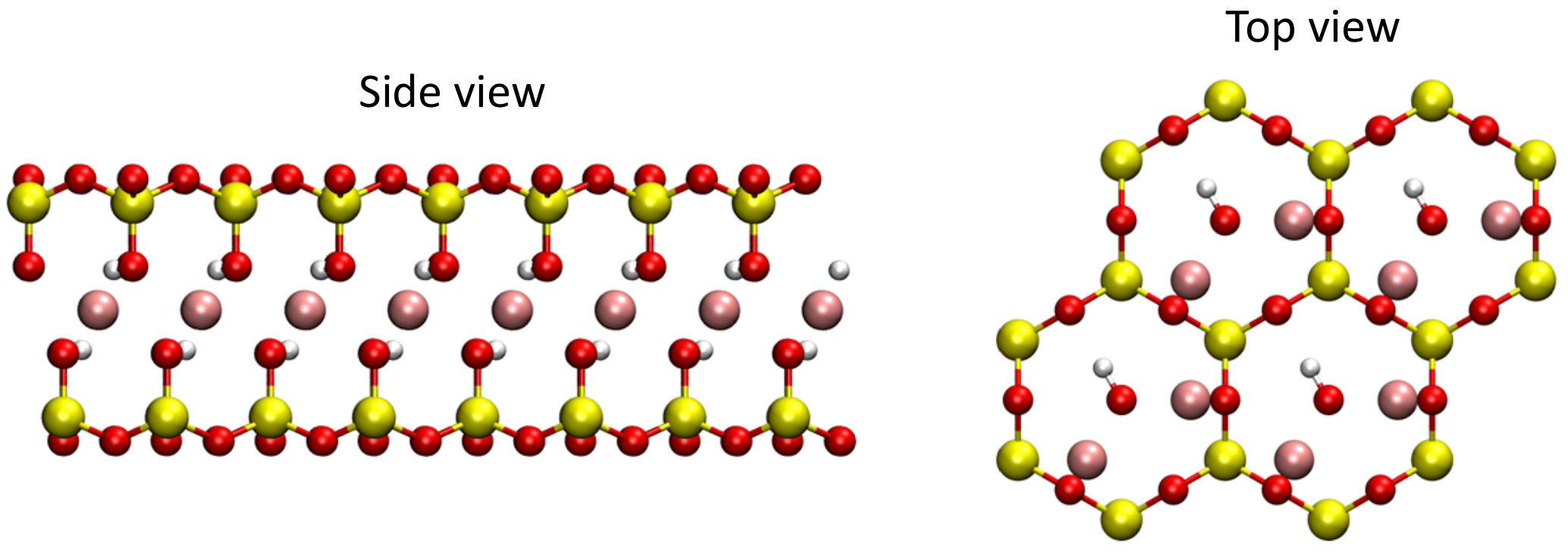}
\par\end{centering}

\protect\caption{Side and top views of a pyrophyllite sheet.\label{fig:Side-and-top} Silicium atoms are in yellow, aluminum in pink, oxygen in red and hydrogen in white.}
\end{figure}

Pyrophyllite is a neutral clay of space group $2/m$, monoclinic,
cleaved along orientation $\left\{ 100\right\} $. It is a stacking
of 9 atomic layers. The top layer is composed of oxygen (O), just
below which are silicium (Si) atoms. Both O and Si each form a two-dimensional
honeycomb structure. Aluminum (Al) atoms compose the central layer,
in 2 thirds of octahedral sites. Between the central layer and the
O and Si layers lies an oxygen-hydrogen (O-H) layer. Their, O atoms
are at the center of the hexagons formed by Si and top O, with the
O--H bond pointing toward the empty octahedral site. Top and side views
of pyrophyllite sheets are given in Fig.~\ref{fig:Side-and-top},
and coordinates of the atomic layers along the axis perpendicular
to the sheet, $z$, are given in Tab.~\ref{tab:Coordinates-of-sheet}.
We have chosen the central layer as $z=0$.

\begin{table}
\begin{center}
\begin{tabular}{|c|c|c|c|c|c|c|c|c|c|}
\hline 
Type & O & Si & O & H & Al & H & O & Si & O\tabularnewline
\hline 
\hline 
$z$~(\AA) & 39.03 & 39.62 & 41.21 & 41.30 & 0.00 & 1.0 & 1.09 & 2.68 & 3.27\tabularnewline
\hline 
\end{tabular}
\par\end{center}

\protect\caption{Coordinates of atomic layers of the pyrophyllite clay along the $z$
axis, perpendicular to the plane of the layers.\label{tab:Coordinates-of-sheet}}
\end{table}

\subsection{Numerical Aspects}
In both MD and MDFT simulations, atoms of the surface interact with
the solvent, water, through optimized Lennard-Jones (van der Waals)
and point charges from the force field for clays, CLAYFF\cite{CLAYFF_cygan_molecular_2004}.
It is a general purpose force field for simulations involving multicomponent
minerals, and their interfaces with a solvent. The CLAYFF force field
for pyrophyllite is described in Tab.~\ref{tab:Description-of-the}.

\begin{table}
\begin{center}
\begin{tabular}{|c|c|c|c|c|}
\hline 
Molecule & Atom & $\epsilon$~(kJ/mol) & $\sigma$~(\AA) & $q$~(e)\tabularnewline
\hline 
\hline 
Water SPC/E & O & 0.65 & 3.165 & $-0.8476$\tabularnewline
\hline 
 & H & 0 & 0 & 0.4238\tabularnewline
\hline 
\hline 
Pyrophyllite & Al & 5.56388e-6 & 4.27120 & 1.575\tabularnewline
\hline 
 & Si & 7.7005e-6 & 3.30203 & 2.1\tabularnewline
\hline 
 & O$_{\text{\text{G}}}$ & 0.65019 & 3.16554 & $-1.050$\tabularnewline
\hline 
 & O$_{\text{H}}$ & 0.65019 & 3.16554 & $-0.950$\tabularnewline
\hline 
 & H$_{\text{G}}$ & 0.0 & 0.0 & 0.425\tabularnewline
\hline 
\end{tabular}
\end{center}
\protect\caption{Force field used to model the pyrophyllite and the water solvent.
Pyrophyllite parameters are extracted from the CLAYFF force field.\label{tab:Description-of-the}}
\end{table}

\subsubsection{MDFT}

The density that minimizes the Helmholtz free energy functional, $\mathcal{F}$,
defined in Eqs.~\ref{eq:FF} to \ref{eq:Fid} is the equilibrium
density in the presence of the external potential. In order to find this minimum numerically, we discretize the molecular density, $\rho\left(\mathbf{r},\boldsymbol{\Omega}\right)$,
onto a space and an angular grid. The tridimensional position grid is
orthorhombic and has typically 3 to 4 points per \AA. The angular
grid is of Lebedev type for the two Euler angles describing the spherical
orientation of the molecular main axis ($C_{2}$ for water). It is
a regular, Gauss-Chebyshev quadrature, for the third Euler angle describing
the molecular rotation around this main axis. Large numbers of fast
Fourier transformed are used to handle convolutions, and this critical
part of the algorithm is executed by the Fastest Fourier Transform
in the West 3 library (FFTW3). As an optimizer, we use the Limited-Memory
Broyden-Fletcher-Goldfarb-Shanno (L-BFGS) quasi-Newton method, as
implemented by Byrd, Lu, Nocedal and Zhu\cite{LBFGS_byrd_limited_1995,LBFGS_zhu_algorithm_1997}.
Its main advantage is to only require $\mathcal{F}$ and its first
functional derivative with respect to the density field. The second
order derivatives, often needed in Newton-type algorithms, is here
approximated with the first derivatives of past iterations. This is
a critical advantage for our minimization process. 

The external potential, $\phi_{n}\left(\mathbf{r}\right)$, and the
electric field $\mathbf{E}\left(\mathbf{r}\right)$ are computed only
once, in the earliest steps of the simulation, before the minimization process.
We use safe cut-offs of 4~$\sigma_{i}$ for the Lennard-Jones interactions.
The electric field is computed by first extrapolating the solute charge
density on the grids and then solving the resulting Poisson equation
in Fourier space.

\begin{figure}
\begin{centering}
\includegraphics[width=0.9\columnwidth]{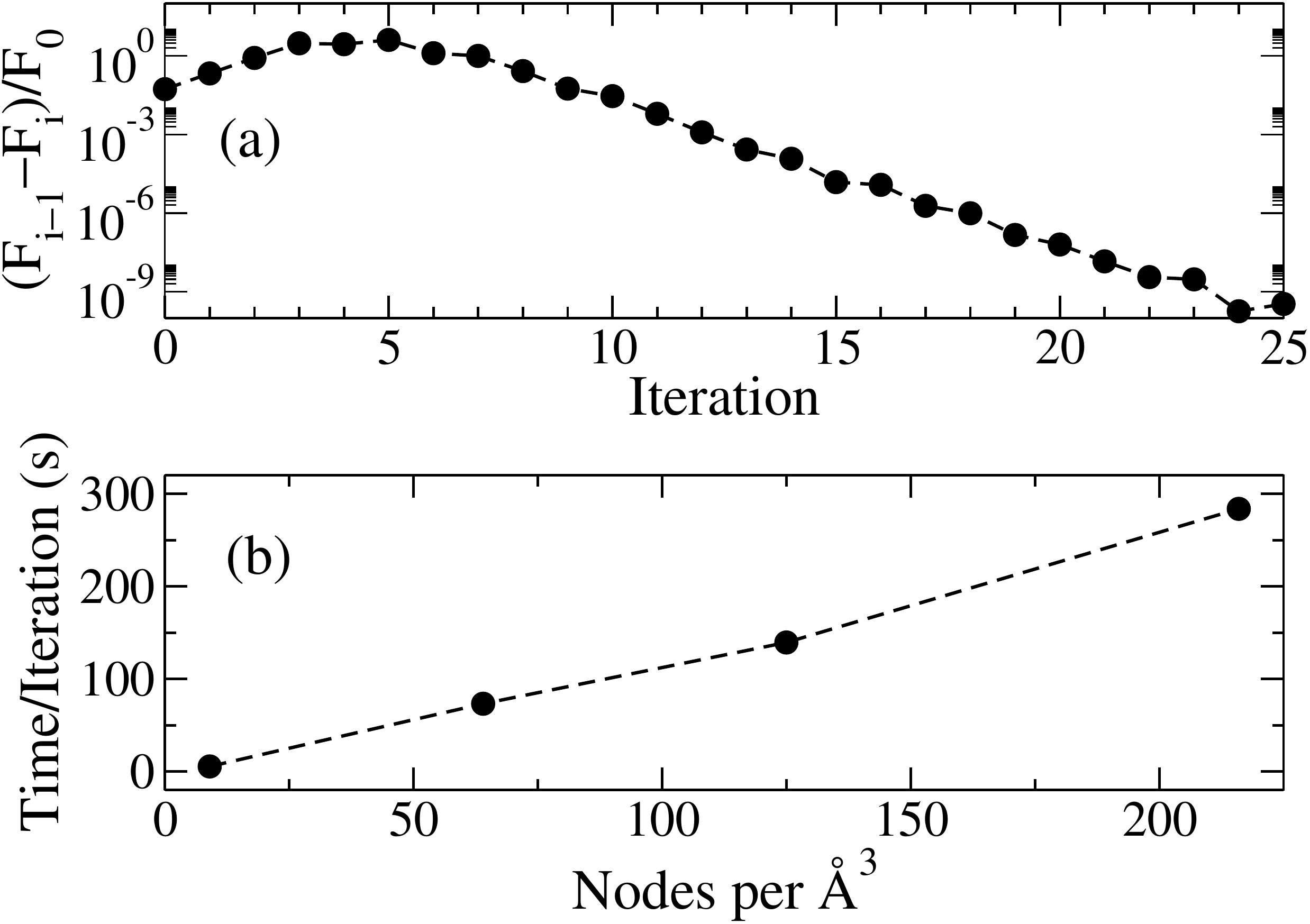}
\par\end{centering}

\protect\caption{(a) Difference in the solvation free energy functional $F$ between
two iterations normalized by the initial value $F_{0}$, during the
iteration process. The grid is composed of $4\times4\times4$ nodes
per \AA$^{3}$ and 24 discrete orientations of the water molecules
per node. A typical minimization is performed with a convergence criteria
of $10^{-5}$ relative variation in free energy, usually reached in
15 iterations. (b) CPU time in seconds per iteration for grid
meshes of 3, 4, 5 and 6 nodes per \AA~and 24 molecular orientations per node. The solvated solute consists
in the pyrophylite sheet described in Section~\ref{Sys_def}. The supercell is approximately 63~nm$^{3}$.\label{fig:numerics}}

\end{figure}

In this work, we typically used 64 position nodes per \AA$^{3}$
and $\approx30$~Euler angles per node, for a total of almost 2000~variables
to be optimized per \AA$^{3}$ by L-BFGS. The convergence is reached
in approximately 15 iterations, for a convergence criteria of $10^{-5}$,
as illustrated in Fig.~\ref{fig:numerics}. The whole minimization
process lasts approximately 25~min on an ordinary laptop, without
parallelism. This is \emph{three order of magnitude faster }than reference
molecular dynamics.

\subsubsection{Molecular Dynamics}

The simulation box contains 1280 clay atoms, for a dimension $L_{x}\times L_{y}\times L_{z}=41.44\times35.88\times42.30$~\AA$^{3}$,
which corresponds to 64 elementary, monoclinic, unit cells of formula
Al$_{2}${[}Si$_{4}$O$_{10}${]}(OH)$_{2}$. The supercell is thus
approximately 63~nm$^{3}$. 
While DFT corresponds to the grand-canonical ensemble, molecular simulations were performed in the canonical ensemble, under conditions such that water far from the surfaces does not feel the effect of the latter. This is achieved in practice by performing prior Monte-Carlo simulations in the NPT ensemble for increasing water contents until the equilibrium distance between the surfaces is sufficient to ensure that the mid-plane density is uniform and equal to the density of bulk water. This ensure to set the solvent chemical potential to the value used in MDFT simulation.
 The obtained configuration has been used 
as the starting configuration for the molecular dynamics simulation.
When simulated by all-atoms molecular dynamics, 1600
water molecules are necessary to fill the pore  
of pyrophyllite sheets, separated by distance $L_{z}$ at 300~K.
The temperature is controlled via a Nos\'e-Hoover thermostat.
The simulation package used here is LAMMPS ("Large-scale Atomic/Molecular Massively Parallel Simulator")\cite{plimpton_fast_1995}.
 After a phase of equilibration, all structural quantities were collected and averaged on a 8~ns trajectory with a 1~fs timestep. Because
periodic boundary conditions are applied, the physical system being simulated by MD and MDFT is thus an infinite stacking of infinite sheets.

The local water density, $n\left(\mathbf{r}\right)$, is calculated
as the time averaged density in an elementary volume of $1.0^{}$~\AA$^{3}$.
The densities in the planes perpendicular to the clay sheet are averaged
in the $z$ planes in an elementary volume of $0.3\times L_{x}\times L_{y}$~\AA$^{3}$,
where $L_{x}$ and $L_{y}$ are the length of the supercell in the
$x$ and $y$ directions. The spatial densities
of the average dipoles were calculated in the same way.

\section{Results}\label{sec:Results}
In a previous paper, we have shown that MDFT allows the study of the solvation of complex systems (pyrophyllite, again). This was done
with the most simple dipolar solvent: the Stockmayer fluid, a linear molecule modeled by a unique central Lennard-Jones potential associated to two charges separated by a a distance that
induces the same dipole as water. The functional we used in this previous paper was thus at the dipolar level only.
Indeed, as expressed in the previous paper, we did not want to illustrate the difficulties linked to finding the most appropriate functional for water, which is
in itself a long term goal, but we wanted to show the efficiency of MDFT in a well-known framework.
Here the water molecules can rotate around their $C_2$ axis.
Regarding the functional, it is developed to a fully multipolar level.
 We first compare the predictions of the molecular
density functional theory for the solvent density with our reference
all-atom simulations that give exact results. Then, energetics are discussed. We finally analyze
the role of electrostatic interactions exerted by the surface on the
solvent on these structural and energetic properties.

\subsection{Density Profiles}

\begin{figure} 
\begin{centering}
\includegraphics[width=0.9\columnwidth]{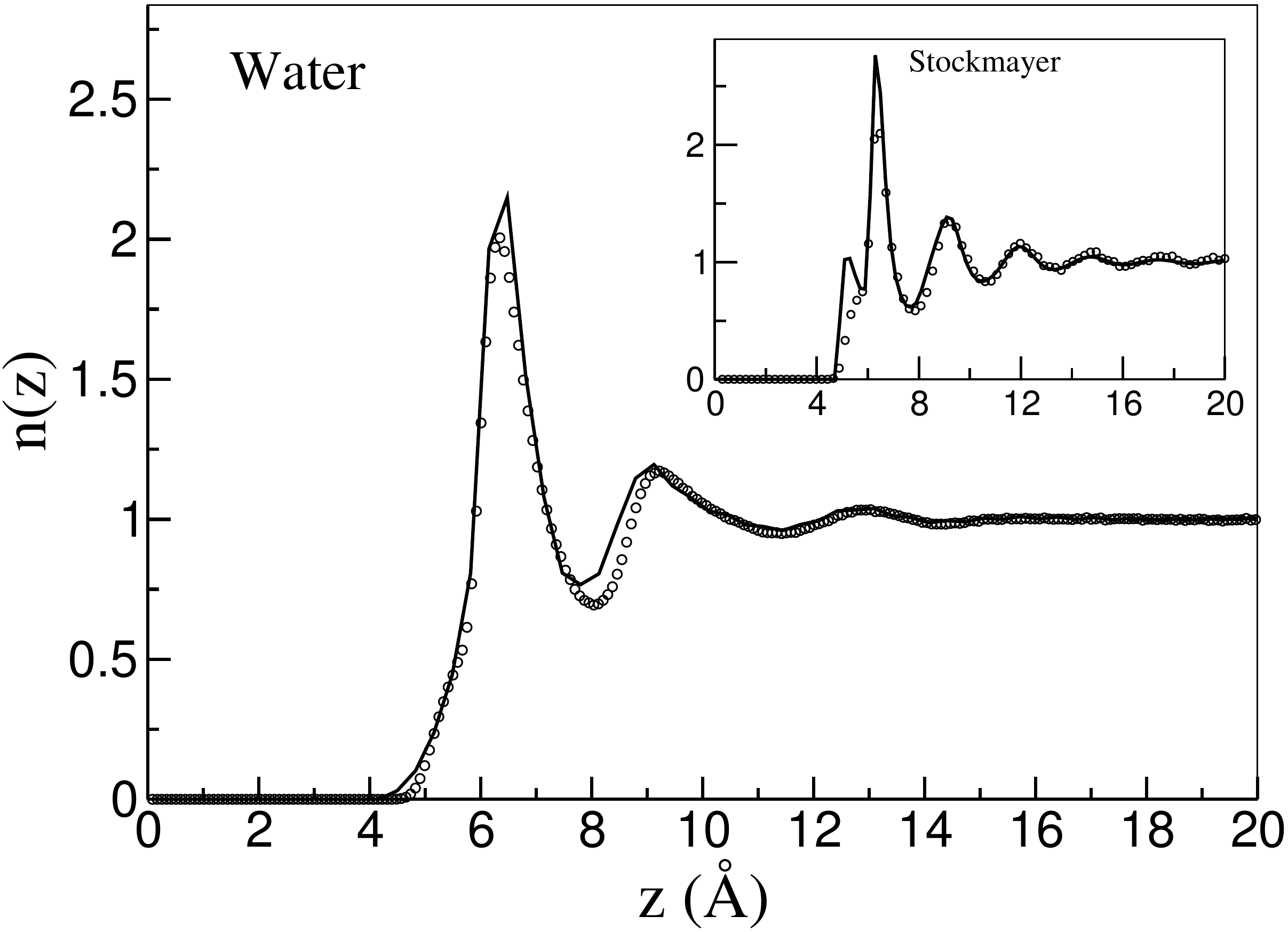} 
\par\end{centering}

\protect\caption{Average density profile of water in the direction $z$ perpendicular to the
pyrophyllite sheet. See Eq.~\ref{eq:n_z} for the definition of $n_{z}\left(z\right)$.
The profile of reference, calculated by molecular dynamics, is plotted
as open circles. MDFT results are shown as a plain lines. The same quantities obtained with the Stockmayer fluid \cite{levesque_solvation_2012} are displayed in  the inset.\label{fig:Density-profile.}}
\end{figure}

The first observable to be computed in a slit pore like the interface
of interest here is the averaged number density profile in the direction
$z$ perpendicular to the clay sheet:
\begin{equation}
n_{z}\left(z\right)=\frac{1}{L_{x}L_{y}}\iint\frac{n\left(\mathbf{r}\right)}{n_{b}}\text{d}x\text{d}y,\label{eq:n_z}
\end{equation}
where the pyrophyllite sheet is parallel to plane $(x,y)$,
and the system has size $L_{x}\times L_{y}\times L_{z}$. Both MD
and MDFT density profiles are given in Fig.~\ref{fig:Density-profile.}, where the results obtained
 for the Stockmayer fluid in Ref.~\cite{levesque_solvation_2012} are also plotted.

Importantly, there is a quantitative agreement between the profiles
calculated with MDFT and the reference profile.

Three peaks are observed, followed by an homogeneous bulk number density,
$n(\mathbf{r})/n_{b}=1$. The first peak is broad and high. It is
found at $z=6.3$~\AA, with $n_{z}=2.2$. This peak shows a small
shoulder at $z\approx5.7$~\AA. This shoulder should be reminded later
in this work. The first peak corresponds to the first solvation layer
of the clay. The second solvation sheet is of relatively small height
($n_{z}\approx1.2$). It is found 2.9~\AA~farther from the clay.
A notably depleted zone is found in between the water layers we have
just described. An echo of the second water layer is found at $z=13.0$~\AA.
Its height is almost negligible, $1.03$, and would remain unseen
without the also almost negligible depletion the precedes it. At
higher $z$ coordinates, the solvent is homogeneous, \textit{i.e.}, is at bulk
density $n_{b}\approx0.033$~molecule per \AA$^{3}$.

All these structural properties are captured by both molecular dynamics
and molecular density functional theory, in quantitative agreement,
and with the same level of detail. Here, we recover the good agreement between MD and MDFT we found previously for the stockmayer fluid\cite{levesque_solvation_2012}, even if the solvent and the functional are much more complicated. It is worth mentioning that when we consider a purely dipolar fluid, we overestimate the prepeak with MDFT. This effect is corrected with better description of the solvent.
 Nevertheless, by averaging the
density $n\left(\mathbf{r}\right)$ in the $\left(x,y\right)$ plane,
one does not figure out the real \emph{local }water structure on top
of the surface.

\begin{figure}
\begin{centering}
\includegraphics[width=7.5cm]{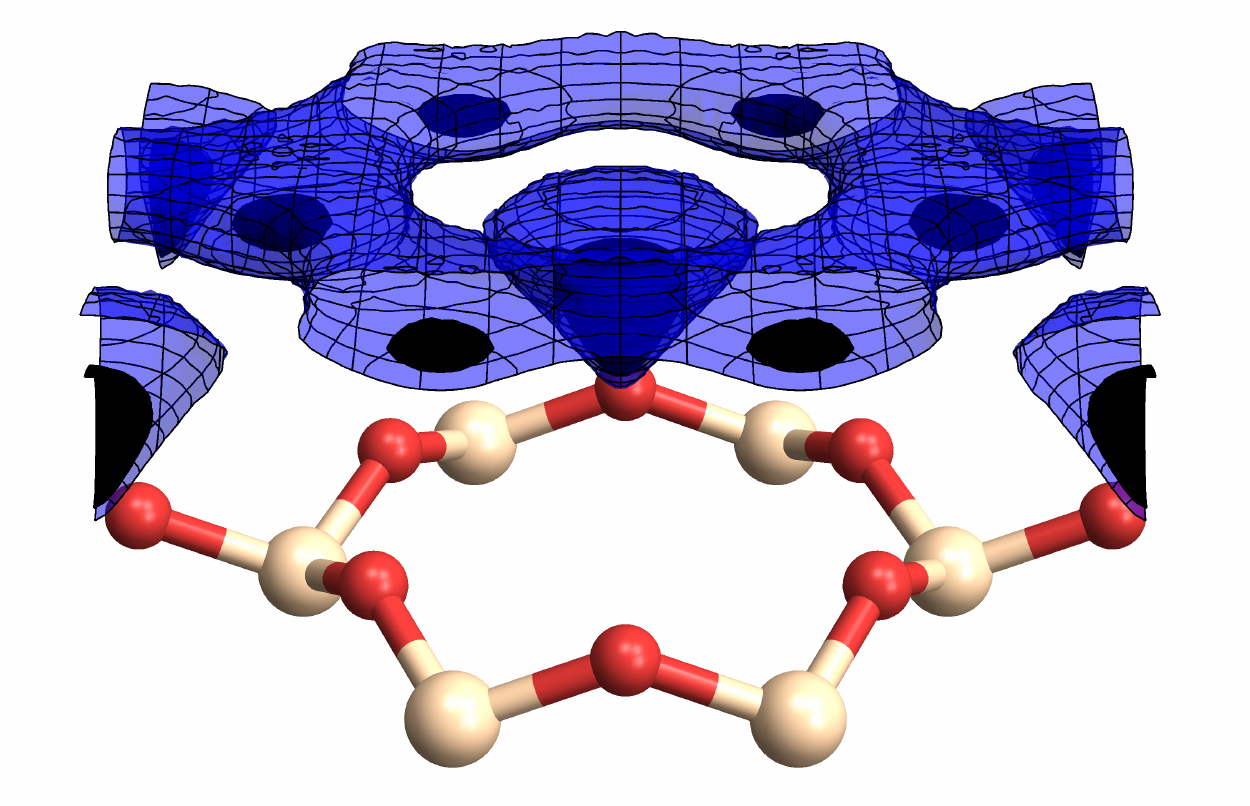}
\par\end{centering}

\protect\caption{Isosurfaces of density $n\left(\mathbf{r}\right)=2$ (in blue) and
4 (in black). The honeycomb unit cell is represented for clarity,
with oxygen and silicium atoms in red and white, respectively.\label{fig:Isosurfaces-of-density}}
\end{figure}

It is worth emphasizing here that extracting \emph{local }information\emph{
}from MD and MC simulations is time-consuming. Indeed, because of
the intrinsic nature of these sampling methods, one cannot
compute exactly a local quantity like $n\left(\mathbf{r}\right)$,
but rather its convolution by some elementary volume. This means that
one effectively measures $n_{z}\left(z\right)\ast\Delta z$ or $n\left(\mathbf{r}\right)\ast\Delta V$,
where $\Delta z$ and $\Delta V$ are elementary dimensions chosen
arbitrarily and $\ast$ denotes the convolution, not $n_{z}\left(z\right)$ or $n\left(\mathbf{r}\right)$.
If the convolution kernels $\Delta z$ and $\Delta V$ are too small,
the time needed to extract with good statistics the maps shown in Fig.~\ref{fig:Isosurfaces-of-density},
for instance, is increased to a critical point. This problem is \emph{not
}relevant in DFT calculations, because it is \emph{not }a method relying
on sampling. Indeed, and for the best here, the natural variable in
MDFT is the local one-particle density, $n\left(\mathbf{r}\right)$.
As an illustration, and in order to get insight to the three dimensional
solvation structure of water on top of the clay, we use MDFT. In Fig.~\ref{fig:Isosurfaces-of-density},
we show the tridimensional contour plots of the isosurface of high
number densities $n\left(\mathbf{r}\right)=2$ and 4. The highest
densities are found on top of the center of the hexagons, and of silicium
atoms of the top-layer (this region is responsible of the prepreak in Fig.~\ref{fig:Density-profile.}). Then, a zone of high density forms on top
of the hexagone (theses regions are responsible of the main peak in Fig.~\ref{fig:Density-profile.}).

\begin{figure}
\begin{centering}
\includegraphics[width=1\columnwidth]{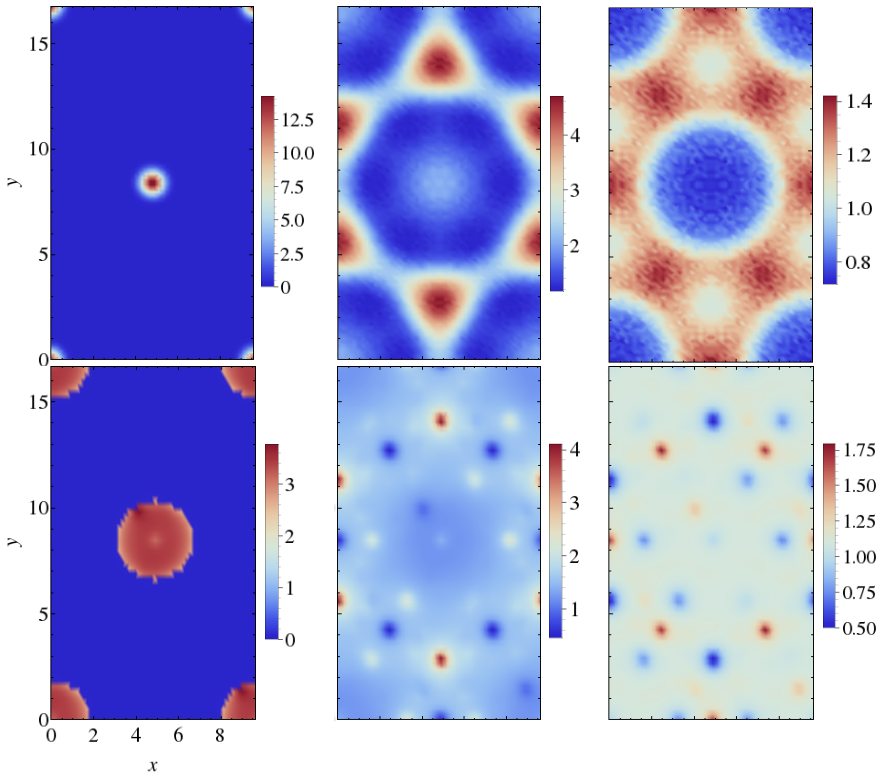}
\par\end{centering}

\protect\caption{Maps of the normalized number density $n\left(\mathbf{r}\right)/n_{b}$
of water in the planes of the prepeak (left), main peak (center) and
secondary peak (right), as calculated by molecular dynamics (top)
and molecular density functional theory for water (bottom). \label{fig:density-maps}}
\end{figure}

Once again, in order to compare MDFT and reference results more locally,
we plot in Fig.~\ref{fig:density-maps}, the normalized density
maps in the planes defined by $z=5.7$, 6.4 and 9.2~\AA~corresponding
to planes of the shoulder, the main and secondary peaks in the density
profile of Fig.~\ref{fig:Density-profile.}. In the shoulder, the
high density zone is strongly localized in the center of the hexagons,
really close to the hydrophobic surface. At such close distance from
the clay, water is repelled from everywhere but from this tiny zone.
Such localization induces that during the averaging leading to Fig.~\ref{fig:Density-profile.},
only a shoulder is found. MDFT underestimate the density there, but
overestimates the width.

In the first solvation layer, \textit{i.e}., seen as the main peak of the density
profile, water molecules are found on top of the Si atoms. There are
2.99~\AA~in between water molecules. This property is found in both
MD and MDFT. 

The second solvation layer, \textit{i.e.}, the secondary peak of the density
profile, shows high density zones on top of the oxygens and depleted
zones on top of Si atoms. There, the variations in number density
are not as important as in the previous peaks, but steel between 0.5
and 1.75.

From Fig.~\ref{fig:density-maps} it is clear that, despite some disparities, the agreement on density prediction between
 MD and MDFT is semi-quantitative. Moreover, one can easily see, and we checked it carefully, that
convoluting the MDFT map by kernel $\Delta V$ of length 1.0~\AA~gives
back smoothed maps close to the MD maps. The better resolution of MDFT cause theses discrepancies
that is why the results are presented in this more meaningful maps, although less comparable.
 Given its numerical efficiency with respect to MD and its semi-quantitative results, MDFT 
 seems to be a method of choice to address this problem.
 

\subsection{Orientational properties}

We now turn to the orientational properties of the water molecules
that solvate the pyrophyllite surface. In molecular dynamics and Monte
Carlo simulations, it is worth emphasizing that orientational properties
are difficult to extract, for the same reason as local properties.
They are indeed even more local in the phase space. The numerical
issue of sufficient sampling of the phase space is recurrent in numerous
simulation techniques where orientational degrees of freedom (from
electronic spins to molecular orientations) induces new and rich behaviors\cite{lavrentiev_magnetic_2010, levesque_simple_2011, levesque_electronic_2012}.
For its part, the density functional theory framework and our MDFT
implementation give a direct access to this quantity through the full
orientational density $\rho\left(\mathbf{r},\boldsymbol{\Omega}\right)$
defined in Eq.~\ref{eq:rho_solvant}. It is even easier to represent
orientational observables thanks to the separation of $\rho$ into
the molecular density field, $n\left(\mathbf{r}\right)$, and the
polarization density field, $\mathbf{P}\left(\mathbf{r}\right)$,
defined in Eq.~\ref{eq:n} and \ref{eq:P}: They are the natural
variables and outputs of MDFT.

\begin{figure}
\begin{centering}
\includegraphics[width=0.9\columnwidth]{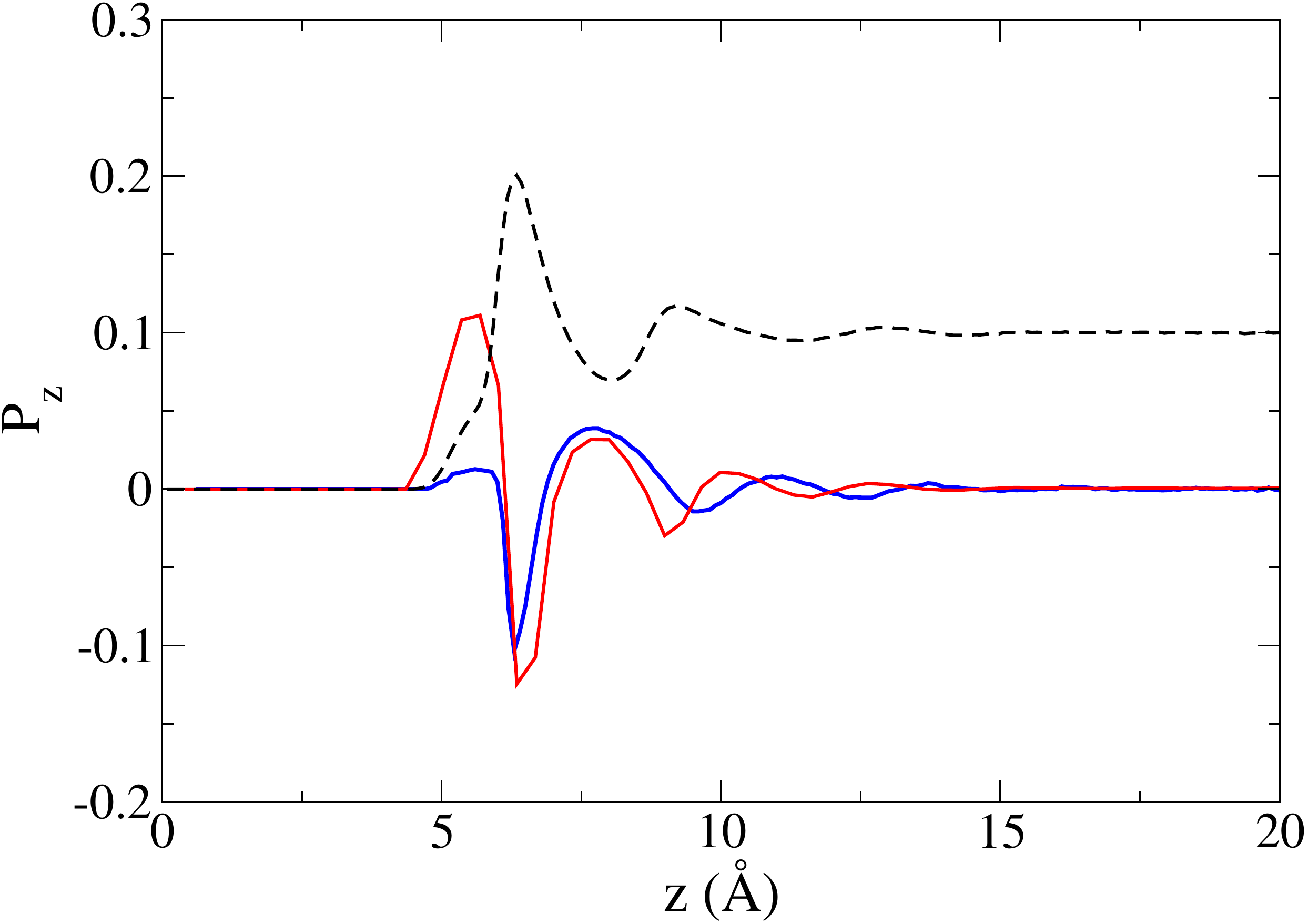}
\par\end{centering}

\protect\caption{Projection of the polarization on the $z$ axis, $P_z$, in the direction $z$ perpendicular to the pyrophyllite sheet. The result obtained by MDFT is displayed in red,
the one by MD is in blue. The density profile, $n_z(z)$, is also shown as a black dashed line, in arbitrary units.
\label{fig:z_Pz_all}}
\end{figure}

In Fig.~\ref{fig:z_Pz_all}, we report the projections of $\mathbf{P}$
on the $z$ axis, noted $P_{z}$,
averaged in each plane $z$, as a function of $z$, as computed by MD and MDFT.
The projections on the $x$ and $y$ axes have also been computed and found negligible: The maximum values of $P_x/P_z$ and $P_y/P_z$ are of the order of $10^{-3}$.
The polarization density is found to be aligned in the direction perpendicular 
to the clay sheet, both in MD and MDFT.

Extrema in polarization density correspond to extrema in water
density. A first positive peak in $P_{z}$ is found at the location of the
shoulder in $n_{z}$. Another negative extremum corresponds to the main peak
in the density profile. The first polarization extremum is smoother and larger than the second one.
 Between each maximum, the sign of $P_{z}$ changes. The intensity of the
 macroscopic polarization is linked to the density of molecules 
 and to their orientation in the region of the space considered.
Note that since the $z$ axis is always oriented in the direction
leaving the surface, a positive $P_{z}$ means that the dipole of
the water molecule, pointing from the oxygen atom to the center of
mass of the hydrogen atoms, points outward the surface, \textit{i.e.}, that
it is the oxygen atom that is the closest to the surface. We can now
see that in the first density zone, \textit{i.e.}, in the shoulder, the oxygen
atoms of the water molecules are globally the closest to the surface.

Water molecules on top of Si atoms,  (in the main peak), point toward the surface, \textit{i.e.}, hydrogens
are globally the closest to the surface. The secondary peak is almost
not polarized, there are no preferential orientation of the water molecules in this high density region.

For the Stockmayer fluid, MD and MDFT results were in quantitative agreement. Here, qualitative agreement is found.
The polarization profiles obtained by MDFT for water solvent is similar to the one obtained with the Stockmayer fluid\cite{levesque_solvation_2012}.

These informations about $P_{z}$ highlighted the layer-by-layer structuration
of the water molecules, but is not enough to conclude on the importance of orientation.
Indeed, a strong polarization can be due either to a small amount of highly oriented molecules or to
a large number of water
molecules, each having some small preference for a given or several
orientations.

\begin{figure}
\begin{centering}
\includegraphics[width=0.9\columnwidth]{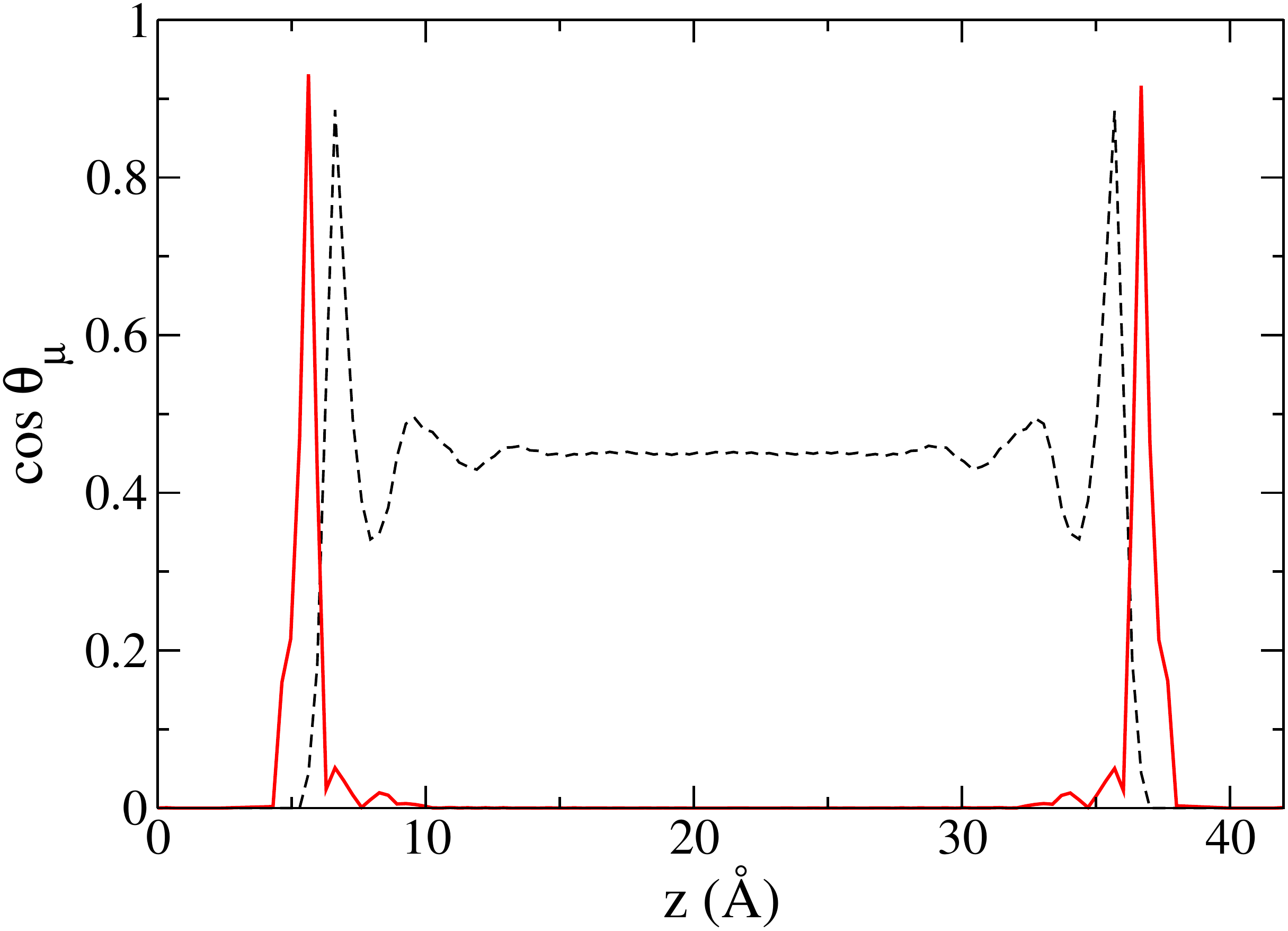}
\par\end{centering}

\protect\caption{The average molecular orientation, $\cos\Theta_{\mu}\left(z\right)$,
along the normal to the surface, $z$, obtained by MDFT, is shown in full red line.
The density profile, $n_{z}\left(z\right)$, is shown in black dashed
line with arbitrary units to ease the understanding of the position
of the peaks.\label{fig:The-average-molecular}}
\end{figure}

To answer this question, we define an orientational order parameters,
the local molecular orientation, $\cos\Theta_{\mu}\left(\mathbf{r}\right)$:
\begin{equation}
\cos\Theta_{\mu}\left(\mathbf{r}\right)=\frac{\left\Vert P_{z}\left(\mathbf{r}\right)\right\Vert }{n\left(\mathbf{r}\right)}.
\end{equation}
The average of the local molecular orientation in the plane $z$,
$\cos\Theta_{\mu}\left(z\right)\equiv\left\langle \cos\Theta_{\mu}\left(\mathbf{r}\right)\right\rangle _{xy},$
is plotted in Fig.~\ref{fig:The-average-molecular} for MDFT results.

It shows that in the shoulder, that corresponds to this highly dense
zone at the center of the hexagons close to the surface, the water molecules are completely oriented.
 We also showed that this zone have a positive $P_{z}$, which means that water molecules
has hydrogen atoms pointing outward the surface. In fact, the oxygen
atom of the water molecule is as close as possible of the positively
charged silicium atoms, which are just under O
atoms of the pyrophyllite. On the opposite the water molecules of the first solvation layer are globally oriented
to make the dipoles point toward the surface, but this preferential orientation is very limited as the
value of $\cos \Theta_{\mu}$ is very weak at the distance corresponding to the first maximum of the density.
The second solvation layer is then stacked,
translated on top of this first layer and with almost no preferential orientation.

MDFT allow to fully uncover the fine structural solvation properties
of pyrophyllite by water.

\subsection{Energetics}

As shown in Eq.~\ref{eq:FF}, the solvation free energy $\mathcal{F}$
is decomposed into physically-grounded components: \textit{(i)} the so-called
ideal part, that accounts for the configurational entropy, \textit{(ii)} the
external part, that accounts for the direct solute-solvent interactions,
and \textit{(iii)} the excess part, that accounts for the solvent-solvent correlations,
that may be split into a radial and a polarization component. During
the minimization process, we can monitor their relative importance.
The equilibrium solvation free energy per surface unit of pyrophyllite,
$\gamma_{\text{solv}}$, is given in Table~\ref{tab:The-different-components}.
It is 332.4~mJ/m$^{2}$, a positive value that reflects the known
hydrophobic nature of pyrophyllite.

\begin{table}

\begin{center}
\begin{tabular}{|c|c|c|c|c|c|}
\hline 
 & $\cal F$ & $\cal F_{\text{ext}}$ & $\cal F_{\text{id}}$ & $\cal F_{\text{exc(rad)}}$ & $\cal F_{\text{exc(ori$^{\circ}$)}}$\tabularnewline
\hline 
\hline 
$\gamma_{\text{solv}}$~(mJ/m$^{2}$) & 332.4 & $-12.8$ & 100.4 & 244.5 & 0.3\tabularnewline
\hline 
\end{tabular}
\end{center}
\protect\caption{The different components of the equilibrium solvation free energy,
per surface unit of pyrophyllite. Positive values of $\gamma_{\text{solv}}$
are meant for hydrophobic surfaces.\label{tab:The-different-components}}
\end{table}

The main contribution to the solvation free energy comes from the
radial solvent-solvent interactions, \textit{i.e.}, from the radial part of
the excess contribution. The ideal part is also of great importance.
Notably, the external and polarization parts do not account much in
$\gamma_{\text{solv}}$ . This must not be understood as a demonstration
that polarization and external potentials do not play a critical contribution
to the solvation process. Indeed, if one thinks about a hard wall
solvated by an ideal gas: the positive infinite external potential
inside the wall is the sole cause that induces the zero density inside
the wall. Then at equilibrium, the external part of the functional,
$\cal F_{\text{ext}}$, is zero, since the density is by nature minimizing
its contribution. 

\subsection{The Role of Electrostatics}
In a previous paper, we tested the role of the electrostatic interactions
between the point charges distributed among atomic sites in the clay
within the CLAYFF force field and the solvent molecules. 
Thanks
to the computational efficiency of our MDFT, we were able to systematically
study the effect of scaling the CLAYFF point charges from their original
values to 0. With the much more realistic picture of water as a solvent,
we now do this same numerical experiment. 

\begin{figure}
\begin{centering}
\includegraphics[width=0.9\columnwidth]{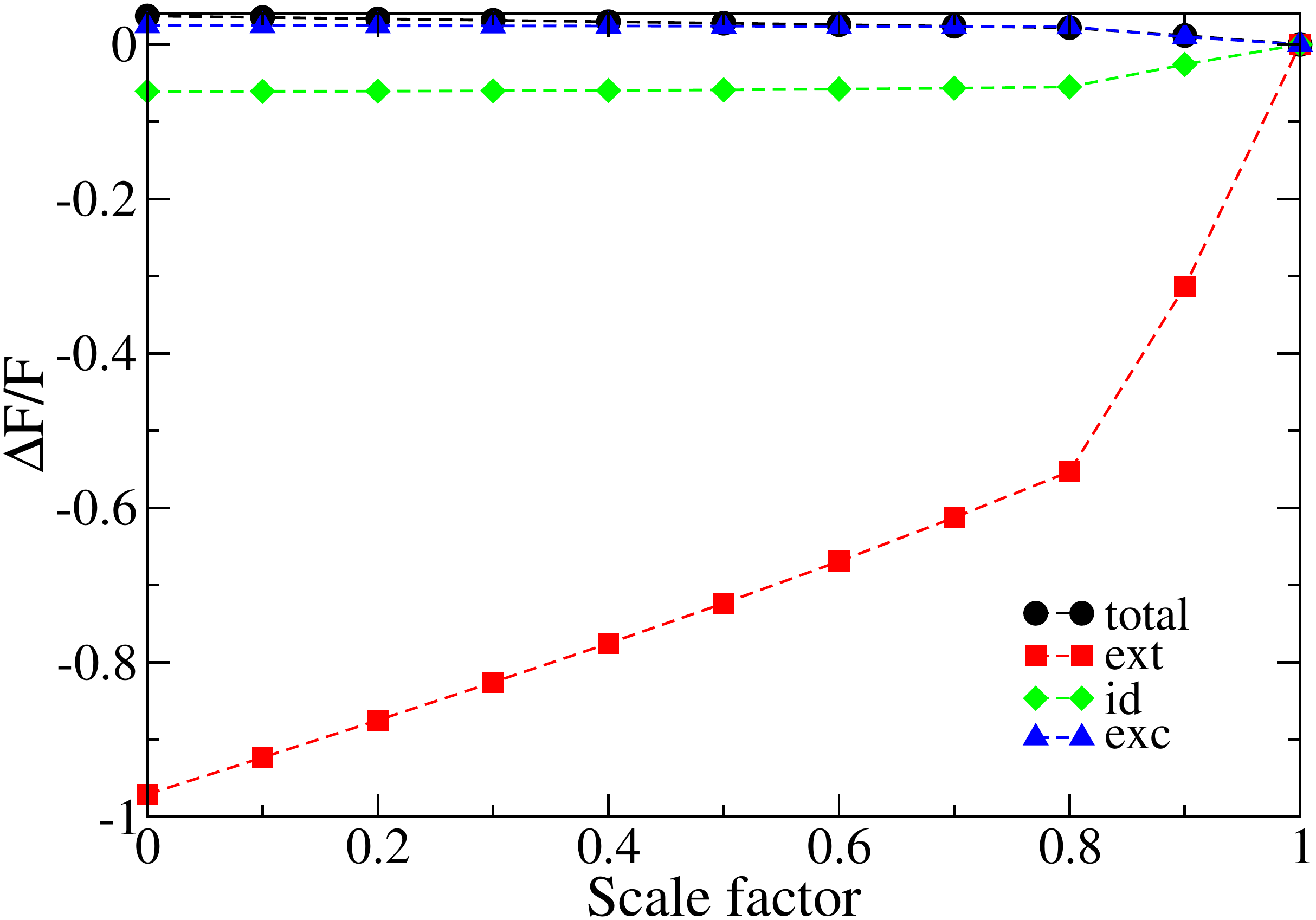}
\par\end{centering}

\protect\caption{Relative evolution of the various parts of the equilibrium solvation
free energy of the clay surface as a function of the solute charge
scale factor. For a scale factor of 0, the charges are turned off.
It is 1 for the charges directly as proposed by CLAYFF. The relative
change in total solvation free energy is 3~\%.\label{fig:Relative-evolution-of}}
\end{figure}

The evolution of the hydration free energy with the scale factor is
given in Fig.~\ref{fig:Relative-evolution-of}. The evolution of
the various components of the solvation free energy is also given.
The solvation free energy does not change by more than 3~\% when
the solute charges are turned off. Furthermore, the density profiles
are not modified by this process, to a point where plots of $n_{z}\left(z\right)$
with and without solute charges are indistinguishable.

When the charges are turned off, the total, radial excess and ideal
part are almost not changing. The external contribution, on its side,
has strong relative changes, but stays small in absolute value. We
have confirmation here that the electrostatic contribution of the clay is not important
for this system, since turning solute charges off does not change
neither the total solvation free energy, nor the solvation structure.
Note that for clay minerals bearing a net total charge (which are in fact more common than electrically neutral ones such as pyrophyllite considered here) with their compensating counter-ions, the effect of polarization could be much more important.

\section{Conclusion} \label{sec:Conclusion}

We have shown that the most recent developments of molecular density
functional theory for water allow to study hydration of a complex
surface of more than a thousand atoms. As an illustration we studied the pyrophyllite clay.
 Structural and orientational hydration properties are quantitatively
comparable to reference all-atoms simulations, while decreasing the
numerical cost by two to three orders of magnitude. The efÞciency of
MDFT is illustrated by extracting Þne local structural
information in half an hour which otherwise requires
several hundred hours of CPU-time with MD. This is due to the intrinsic nature of MDFT,
that relies on the local number and polarization densities as natural
variables.

We were able to investigate how water solvates the pyrophyllite clay,
with (\emph{i}) a single water molecule on top of the hexagone formed
by silicium and oxygen atoms, as close as possible of the plane carrying
Si. This single H$_{2}$O is strongly oriented with dipole pointing
outward from the surface; (\emph{ii}) then, water molecules form a layer on top of
Si atoms and are globally oriented to make their dipoles slightly point toward the surface.

The numerical efficiency of our MDFT approach made it possible to
realize a systematic study to analyze the relative contributions of
electrostatic and van der Waals interactions, by slowly turning off
the charges carried out by the surface. We showed that the density
profiles and solvation free energy are not sensitive to the charges
of the clay. This may imply that the role of electrostatics in charged
clays may be reasonably reduced to their charged defects. This means
that the large amount of work dedicated to force field fitting in
this field may perhaps be reduced by a focus on important contributions.
Such work would rely on a systematic parameters study for which MDFT
seems a highly promising tool. This should be done in a future work.

\bibliography{text}

\end{document}